\documentclass{article}

\usepackage{url}
\usepackage{graphicx}
\usepackage{amsmath,amsfonts,amssymb,subfigure}
\usepackage{color}
\usepackage{multirow,}
\usepackage{colortbl}
\definecolor{grisClair}{gray}{0.75}

\title{Analytical Model of TCP Relentless Congestion Control}

\author{R\'emi Diana and Emmanuel Lochin\\
~\\
Universit\'e de Toulouse - ISAE\\
\small \texttt{firstname.lastname@isae.fr}}\normalsize

\date{}

\begin{document}

\maketitle

\begin{abstract}
We introduce a model of the Relentless Congestion Control proposed by Matt Mathis. 
Relentless Congestion Control (RCC) is a modification of the AIMD (Additive Increase Multiplicative Decrease) congestion control which consists in decreasing the TCP 
congestion window by the number of lost segments instead of halving it. Despite some on-going discussions at the ICCRG IRTF-group, this congestion control has, to the best of our knowledge, never been modeled. 
In this paper, we provide an analytical model of this novel congestion control and propose an implementation of RCC for the commonly-used network simulator ns-2. 
We also improve RCC with the addition of a loss retransmission detection scheme (based on SACK+) to prevent RTO caused by a loss of a
retransmission and called this new version RCC+.
The proposed models describe both the original RCC algorithm and RCC+ improvement and would allow to better assess the impact of this new congestion control scheme over the network traffic.
\end{abstract}

\section{Introduction}

Relentless Congestion Control (RCC) is a proposal from Matt Mathis which consists in a simple modification of the
AIMD (Additive Increase Multiplicative Decrease) congestion control algorithm \cite{RCC}. Basically, instead of halving the TCP congestion window after a loss, 
RCC decreases the current congestion window by the number of lost segments. This behaviour can be modeled as a
strict implementation of van Jacobson's Packet Conservation Principle (this principle suggests that a new packet should 
not be placed into the network until an old packet leaves).  
Indeed, during recovery, new segments are injected into the network in exact accordance with those that 
have been delivered to the receiver \cite{RCC-draft}.

RCC is not an AIMD-friendly protocol and as a result, requires that the network allocates capacity through Fair Queuing or Fair Dropping queue management \cite{fairqueuing}. RCC could perform efficiently over network architectures that enable Quality of Service (QoS) guarantees such as \cite{idms01}. Indeed, a decade of research in QoS has shown that the standard TCP reaction to congestion events (which is to halve its congestion window) can be counterproductive over these QoS networks \cite{comcom07}, \cite{analof09}.
The principle to decrease the current congestion window by the number of lost segments can also be implemented inside current congestion controls such as CUBIC, Newreno or Compound. Intuitively, RCC might enhance the performance of standard congestion controls when losses are not due to congestion (over very noisy wireless links) or large bandwidth-delay product networks (LBDP). In the first case, RCC would prevent large congestion window decrease due to error link losses while in the context of LBDP networks and long delay links, RCC might achieve a higher throughput. However, whatever the context of use, there is no existing analytical model of RCC allowing to estimate the expected rate that would achieve this protocol over a given network. In order to clearly assess the impact of a deployment of such TCP modification, it is obviously essential to model this TCP variant.

In this paper we introduce two models. The first one presented in Section \ref{sec:model} does not take RTO into account. Our second model described in section \ref{sec:modelRTO} is an extension that integrates the RTO effects on congestion window evolution. To prevent RTO triggering, we combine RCC algorithm with a lost retransmission detection scheme based on SACK+ \cite{sackplus}. Thus, we implemented in ns-2 an improvement of RCC algorithm that we called RCC+ corresponding to our first model. Our second model corresponds to the original RCC algorithm that do not avoid RTO (also implemented inside ns-2). We present RCC+ in Section \ref{sec:rcc+} and evaluate the accuracy of both models in Section \ref{simulation} with ns-2 simulations. Finally, we conclude this work in Section \ref{sec:conclusion}.

\section{Analytical model}
\label{sec:model}

\begin{figure}[h] 
		\begin{center}	
		\includegraphics[width=8cm]{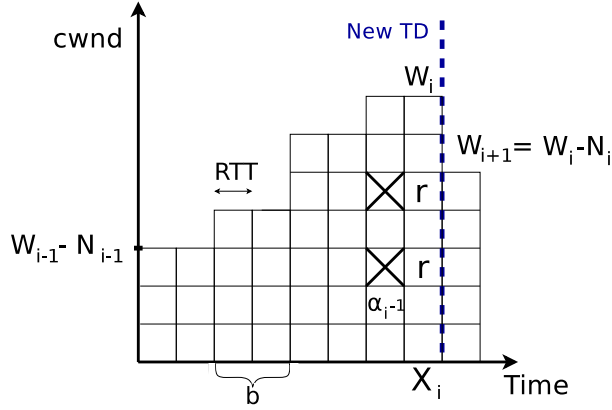} 
		\caption{Evolution of the congestion window size over time.} 
		\label{model_evo_fenetre} 
		\end{center}
\end{figure}

In this section, we develop a stochastic model of TCP Relentless Congestion Control algorithm coupled with the algorithm of selective acknowledgement (SACK).
This leads to a simple analytic expression for the throughput of a TCP Relentless sender as a function of loss rate $p$ and the 
average round trip time ($RTT$). 
Our RCC model is built on the well-known Padhye \textit{et al.} TCP model \cite{Modeling} from which we borrow the notations and the scheme given in Fig. \ref{model_evo_fenetre} to ease the understanding. 

The notations are presented in Fig. \ref{model_evo_fenetre}. The period denoted $TD$ defines an elementary cycle (corresponding to $TDP$ in \cite{Modeling}), delimited by two consecutive decreases of the window $W_i$ where $i$ refers to the $i^{th}$ $TD$.

In the model, we adopt the following rule for all random variables: $V_i$ corresponds to the value of the random variable $V$ in $TD_i$ and its expected value computed on all $TD_i$ is noted $E[V]$.

Let $Y_i$ be the number of packets sent in $TD_i$; $\alpha_i$ the index of the first lost packet; $\beta_i$ the amount of packets sent after $\alpha_i$ to complete the round (corresponding to an $RTT$) and $N_i$ the number of retransmissions done in $TD_i$.
We also define $X_i$ the number of rounds in $TD_i$ and $\frac{1}{b}$ the acknowledgement generating frequency. Thus:
\begin{equation}
Y_i=\alpha_i + \beta_i + W_i
\end{equation}

and

\begin{equation}
E[Y] = E[\alpha] + E[\beta] + E[W]
\label{eqn_y1}
\end{equation}

We now have to derive $E[\alpha]$ and $E[\beta]$. If we consider uniform and independent losses, $\alpha_i=k$ means that the first $k-1$ packets are successfully sent and the $k^{th}$ packet is lost. The packet loss probability is denoted $p$. We can compute $E[\alpha]$ as follows: 

$P(\alpha=k)=(1-p)^{k-1}p$ $\longrightarrow$ $E[\alpha]=\sum_{k=1}^{\infty} (1-p)^{k-1}p.k$
\begin{equation}
	 E[\alpha]= \frac{1}{p}
	 \label{e_alpha}
\end{equation}

$\beta_i$ evolves from $1$, if the losses occur at the end of the window, to $W_i - \frac{1}{b}$, if losses occur at the beginning of the window. 
The uniform aspect of losses implies the uniform distribution of $\beta_i$ in $[1,W_i - \frac{1}{b}]$. It follows that:

\begin{equation}
E[\beta]= \frac{W_i+1-\frac{1}{b}}{2} \simeq \frac{E[W]+1-\frac{1}{b}}{2}
\label{eqn_beta}
\end{equation}

The relation between window size in $TD_{i-1}$ and $TD_i$ can be written as follows: 

\begin{equation}
W_i = W_{i-1} - N_{i-1} + \frac{X_i-1}{b}
\end{equation}

as a consequence: 

\begin{equation}
E[N] = \frac{E[X]-1}{b}
\label{eqn_N}
\end{equation}

The expected value of $N$ can be evaluated with the same assumptions. Basically, if we consider uniform and independent losses with the elementary probability of $p$, then $N$ follows a binomial law of parameters $p$ and $E[\beta]$ (the mean amount of packets sent after the first loss occurs). Thus, $E[N]=1+p(E[\beta])$ and (\ref{eqn_N}) leads to:

\begin{equation}
E[X]=1+b.E[N] = 1 + b\bigg( 1+p \frac{E[W]+1-\frac{1}{b}}{2}\bigg)
\label{eqn_x}
\end{equation}

The evolution of the window size can also be written using slope $\frac{1}{b}$, which corresponds to the evolution pace of $W$.

\begin{eqnarray}
	Y_i &=& \sum_{k=0}^{X_i-1} \bigg(W_{i-1} - N_{i-1} + \frac{k}{b}\bigg) \nonumber \\ 
		&=& X_i\bigg(W_{i-1} - N_{i-1} - \frac{1}{2b}\bigg) + \frac{X_i^2}{2b} \label{eqn_yi}
\end{eqnarray}

If we now take the mathematical expectation of (\ref{eqn_yi}) assuming for a first approximation that {$X_i$} and {$W_i$} are mutually independent sequence of i.i.d. random variables and with $V[X]$ the variance of $X$ we have: 

\begin{equation}
	E[Y]=E[X]\bigg(E[W]-E[N]-\frac{1}{2b}\bigg) + \frac{E[X]^2}{2b}+\frac{V[X]}{2b} \label{eqn_y2}
\end{equation}

If we combine (\ref{eqn_y2}), (\ref{eqn_y1}), (\ref{e_alpha}), (\ref{eqn_beta}) and (\ref{eqn_x}) we obtain:

\begin{eqnarray}
&\frac{bp(4-p)}{8} E[W]^2 + (b-\frac{1}{2}+\frac{p^2(1-b)-3p}{4})E[W]-\frac{p+1}{p}  \nonumber \\
& + \frac{p(-2b^2+b+1)}{4b}-\frac{b^2 +1}{2b}+\frac{V[X]}{2b} + \frac{p^2(b^2-1)}{8b} = 0 \label{eqn_ew}
\end{eqnarray}

By solving (\ref{eqn_ew}) and keeping only the positive root, we obtain a literal expression of $E[W]$ (we do not provide this long expression which is out of interest here). 
Moreover, the duration of $TD_i$, $A_i$, can be expressed as the number of rounds in $TD_i$, $X_i$, multiplied by the average duration of a round, $RTT$: $A_i = X_i.RTT$. Thus, $E[A]=E[X].RTT$.
Now, if we consider $p$ near zero and combine the expressions of $E[W]$, $E[Y]$ and $E[A]$, we obtain the average throughput of a Relentless flow $T_p$ : 

\begin{equation}
	T_p= \frac{E[Y]}{E[A]}= \frac{MSS}{b.RTT.p} + o\bigg(\frac{1}{p}\bigg)
	\label{expr_debit}
\end{equation}

where $MSS$ is the maximum segment size. Let $C$, be the constant term in (\ref{expr_debit}). We have a general expression for the throughput of a Relentless flow which is: 

\begin{equation}
	T_p= \frac{C.MSS}{RTT.p} + o\bigg(\frac{1}{p}\bigg), \ C=\frac{1}{b}
	\label{expr_debit_cste}
\end{equation}

TCP retransmission time out (RTO) is not considered in this model. 
Nevertheless, by combining RCC and SACK retransmission scheme, we are mainly sensitive to one type of RTO which is the loss of a retransmitted packet. 
If the loss rate is low, this event can occur but can be considered as rare. However, we present in the next section an improvement of RCC allowing to prevent RTO due to the loss of a retransmitted packet. 
As a result, the model developed here allows to correctly fit our current implementation (presented in the following Section \ref{simulation}) and motivates why we do not need to take into consideration the RTO.

\section{Analytical model with RTO}
\label{sec:modelRTO}

\begin{figure}[h] 
		\begin{center}	
		\includegraphics[width=\columnwidth]{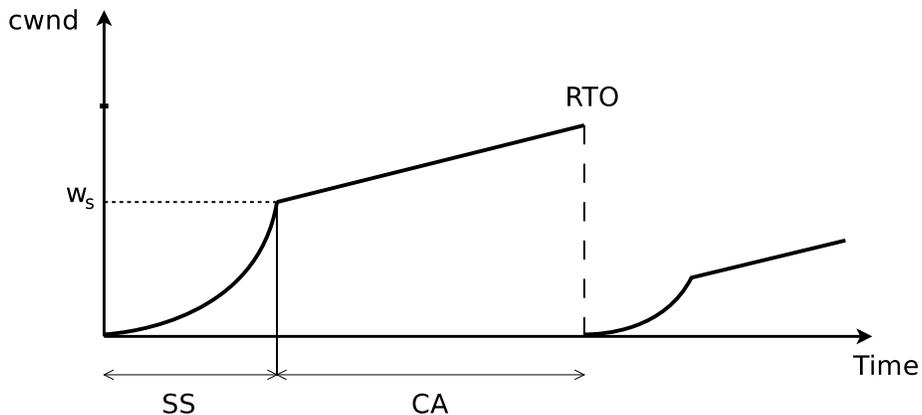} 
		\caption{Evolution of the congestion window size over time in case of RTO.} 
		\label{model_RTO} 
		\end{center}
\end{figure}

In this section, we further develop our previous model to handle the case of RTO. 
We adopt a different approach following the congestion window evolution which behaves as shown in Fig. \ref{model_RTO}.
One of the main challenge of this model is that compared to the standard halving congestion window scheme \cite{Modeling}, we must take into account the number of lost packets during a round. This number, obtained following a probability model, greatly complexifies the model.

The evolution of the congestion window can be described as a repetition of slow start $(SS)$ and congestion avoidance $(CA)$ phases. The $SS$ phase is the same as the one present in TCP Reno or Newreno. This phase is left once the congestion window reaches a threshold or when a loss is detected. In Figure \ref{model_RTO}, $w_s$ represents the value of the congestion window at the end of this phase. The second phase of the cycle corresponds to the phase previously modelled (Section \ref{sec:model}) in which the window is increased by one minus the number of losses after each RTT.

Let $X_{SS}$, $X_{CA}$ be respectively the number of packets sent during the $SS$ and $CA$ phases and $D_{SS}$, $D_{CA}$ the respective duration of $SS$ and $CA$ phases.
Throughout in case of RTO, $T_{RTO}$, can still be expressed as:
\begin{equation}
\label{trto}
 T_{RTO}= \frac{E[X_{SS}]+E[X_{CA}]}{E[D_{SS}]+E[D_{CA}]}
 \end{equation}

In the following, we develop each of the terms of (\ref{trto}):
\begin{equation}
	\label{xss}
	E[X_{SS}]=\sum_{k=1}^{lim} k.P(X_{SS}=k)
\end{equation}
In (\ref{xss}), $lim$ represents that $X_{SS}$ increases until the reach of a threshold or when a loss occurs. As losses are considered to be uniformly distributed, we consider in a first approximation, $lim$ to be in average equals to $ \frac{1}{p} $.

To express the amount of sent packets, we consider the congestion window in the $CA$ phase as an arithmetico-geometric sequence denoted $(W_{CA}^n)_{n \in \mathbb{N}}$. After an RTT, the congestion window is increased by one and decreased by the number of losses which depends on the size of the previous congestion window value. Indeed, as the number of losses follows a binomial law of parameters $p$ and $W_{CA}^n$, the average number of losses is $pW_{CA}^n$. Thus, the first term of this sequence is $w_s$ and the evolution of $W_{CA}$ is given by $ W_{CA}^{n+1} = W_{CA}^n + 1 - pW_{CA}^n$. The general term of this sequence can be written as follows:
\begin{equation}
	\label{wca}
	W_{CA}^{n}(w_s)=(1-p)^n(w_s-\frac{1}{p})+\frac{1}{p}
\end{equation}	

Let $I_{RTO}$ be the index of the RTO event in the $CA$ phase. $I_{RTO}$ evolves from $1$ to $+\infty$.
$E[X_{CA}]$ depends on $w_s$ and is given by:

\begin{eqnarray}
	\label{xca_1}
	E[X_{CA}](w_s) = & \sum_{k=1}^{\infty} \big{[} P(I_{RTO}=k) . \\ \nonumber
		& \big{(}\sum_{j=1}^{k}W_{CA}^j(w_s) - \frac{W_{CA}^k(w_s)}{2}\big{)}\big{]}
\end{eqnarray}

In (\ref{xca_1}), the term $- \frac{W_{CA}^k(w_s)}{2}$ represents the fact that when an RTO occurs, the end of the window is lost. In average, we can consider that the RTO occurs in the middle of the window leading to the loss of the second half of the window.
To trigger an RTO, a packet and its retransmission have to be lost. The probability of this event is $p^2$. RTO can occur for any packet of the window. As a consequence:
\begin{equation}
	\label{prto}
	P(I_{RTO}=k) = p^2W_{CA}^{k-1}(w_s)
\end{equation}

Moreover, as the value of the congestion window is equal to the amount of packets sent during SS $phase$, $w_s$ can evolve from $1$ to $lim$. Thus we have: 

$$P(w_s=k)=(1-p)^{k-1}p$$ 

and (\ref{xca_1}) leads to:

\begin{eqnarray}
\label{xca_2}
	 E[X_{CA}] & =\sum_{i=1}^{lim} \Big{[} p(1-p)^{i-1} \sum_{k=1}^{\infty} \Big{(}p^2W_{CA}^{k-1}(i) \Big{(} \nonumber \\
	 & \Big{(}  \sum_{j=1}^{k}W_{CA}^j(i) \Big{)}- \frac{W_{CA}^k(i)}{2} \Big{)} \Big{)} \Big{]} 
\end{eqnarray}

We now focus on the expression of phases duration. During the $SS$ phase, the amount of sent packets increases as of power of 2.
As a consequence, $E[D_{SS}]$ is given by: 
\begin{eqnarray}
\label{dss}
	E[D_{SS}] &=& \sum_{k=1}^{lim} log_2(k)P(X_{SS}=k) \nonumber \\
			  &=& \sum_{k=1}^{lim} log_2(k)(1-p)^{k-1}p
\end{eqnarray}
and 
\begin{eqnarray}
\label{dca}
	E[D_{CA}](w_s) &=& \sum_{i=1}^{\infty} i P(I_{RTO}=i) \nonumber \\ 
				   &=& \sum_{i=1}^{\infty} i. p^2W_{CA}^{i-1}(ws) 
\end{eqnarray}

As a consequence the global average of $D_{CA}$ is:

\begin{eqnarray}
\label{dca}
	E[D_{CA}] &=& \sum_{j=1}^{lim} P(w_s=j) E[D_{CA}](w_s) \\ \nonumber 
	&=& \sum_{j=1}^{lim} p(1-p)^{j-1} \Big{(} \sum_{i=1}^{\infty} i.p^2 W_{CA}^{i-1}(j)\Big{)}  
\end{eqnarray}

Finally, combining (\ref{trto}), (\ref{xss}), (\ref{prto}), (\ref{xca_2}), (\ref{dss}), (\ref{dca}) we obtain:
\begin{equation}
	\textstyle
	\label{trtofull}
		 T_{RTO} = \frac{\Theta}{\sum_{j=1}^{lim} \Big{(}(p(1-p)^{j-1}) (\log_{2}j + \sum_{i=1}^{\infty} i.p^2 W_{CA}^{i-1}(j))\Big{)} } \\
\end{equation}
 with:
\begin{eqnarray}
		\Theta=&  \sum_{i=1}^{lim} \Big{[} p(1-p)^{i-1}\Big{(}i+\sum_{k=1}^{\infty} \Big{(}p^2W_{CA}^{k-1}(i)  \nonumber\\
		     &   \Big{(}  \Big{(}  \sum_{j=1}^{k}W_{CA}^j(i) \Big{)}- \frac{W_{CA}^k(i)}				{2} \Big{)} \Big{)} \Big{)} \Big{]} \nonumber	         
\end{eqnarray}

Numerically, we observe that $E[X_{CA}](w_s)$ does not depend on $w_s$ leading to the following approximation for $T_{RTO}$:

$$ T_{RTO} \simeq \frac{C_{RTO}MSS}{RTT.p} + o\Big{(}\frac{1}{p}\Big{)} \ \ \ with \ \ C_{RTO} = 0.49 $$ 

We have verified that the last approximation is closed to $T_{RTO}$ and thus considers in the following this latest and simplified expression.

\section{RCC+ in a nutshell}
\label{sec:rcc+}

As basic RCC algorithm with SACK mechanism is not RTO resistant, we propose to implement RCC with SACK+ \cite{sackplus} in ns-2. SACK+ is a SACK extension allowing to prevent RTO due to the loss of retransmitted packets. 

We have implemented an improvement of RCC in order to prevent RTO caused by a loss of a retransmission. Of course, if a packet is lost each time it is retransmitted RTO cannot be avoided. 
For new connection requests, the retransmission timer is initialized to 3 seconds \cite{rfc1122}. By default, a segment with the SYN flag set, is resent no more than three times.
In other words, it must be received before four RTTs elapse, which is approximately the value of the retransmission timer. The configuration of this timer is also possible to enable more retransmission tries. 

In our implementation, we choose a discrete resolution of the lost retransmission problem as in \cite{sackplus}. 
However, we could choose a continuous solution and set a timer to each retransmitted packet. 
We could set this timer to $k*RTT$, with $RTT$ the current estimation of $RTT$ and $k>1$ to prevent spurious retransmissions.

\begin{figure}[htb!] 
\begin{center}
	\includegraphics[width=0.7\columnwidth]{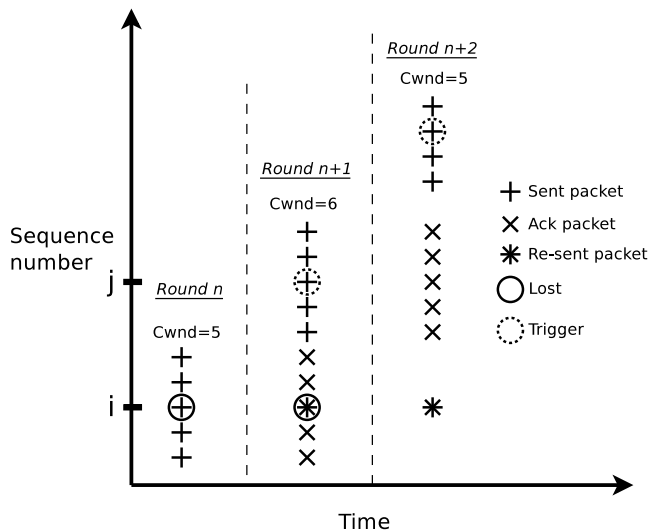}
	\caption{RCC+ behaviour}
	\label{fct_rcc_plus}
\end{center}
\end{figure}

In our case, for each retransmitted packet, we fix a trigger. This trigger corresponds to the acknowledgement of the following \textit{regular} packet which is sent after the retransmission. By \textit{regular}, we refer to packets that do not correspond to a retransmission. Actually, if the regular packet sent after the retransmission is acknowledged before the retransmitted one, we can suppose that the retransmitted packet is lost. As shown in Fig. \ref{fct_rcc_plus}, when packet $i$ is retransmitted in round $(n+1)$, the following regular packet sent which is packet $j$, is used as a trigger for an eventual new retransmission. Indeed, if packet $j$ is acknowledged and not packet $i$, packet $i$ is considered as lost and re-emitted. In this example, as the first retransmission of packet $i$ is lost, packet $j$ is acknowledged and not packet $i$. This leads to a second retransmission of packet $i$ in round $(n+2)$ and the intialisation of a new trigger. Anyway, all these solutions are only different by their implementation. In a general manner, the result remains the same: RTO due to loss of retransmission are avoided.

\section{Implementation and evaluation of RCC and RCC+} 
\label{simulation}
Our first model without RTO consideration corresponds to RCC+ implementation as detailed in Section \ref{sec:rcc+}. The complete model
presented in Section \ref{sec:modelRTO} corresponds to the basic implementation of RCC algorithm.
In this part we first want to demonstrate the accuracy of our models and underline the interest brought by RCC+.

Our models give simple approximations of the throughput of an RCC flow as a function of the $RTT$ and the loss rate $p$. 

\begin{figure}[htb!] 
\begin{center}
		 \subfigure[RCC algorithm]{
			\includegraphics[width=0.90\columnwidth]{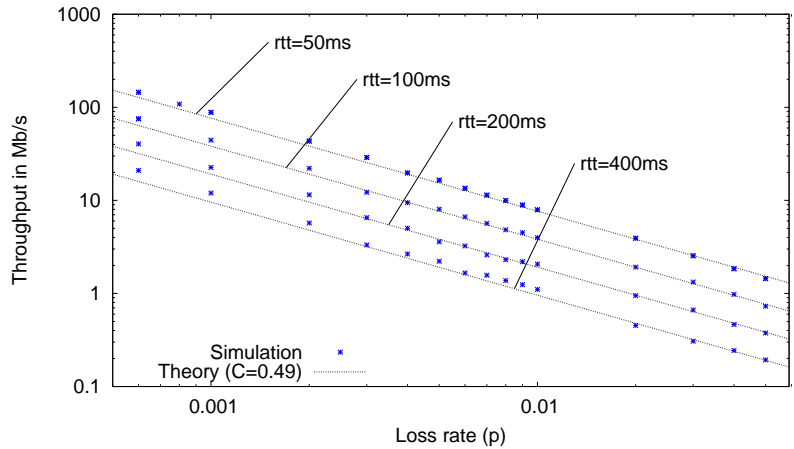}\label{result_simu_rcc_RTO}}
		 \subfigure[RCC+ algorithm]{
			\includegraphics[width=0.90\columnwidth]{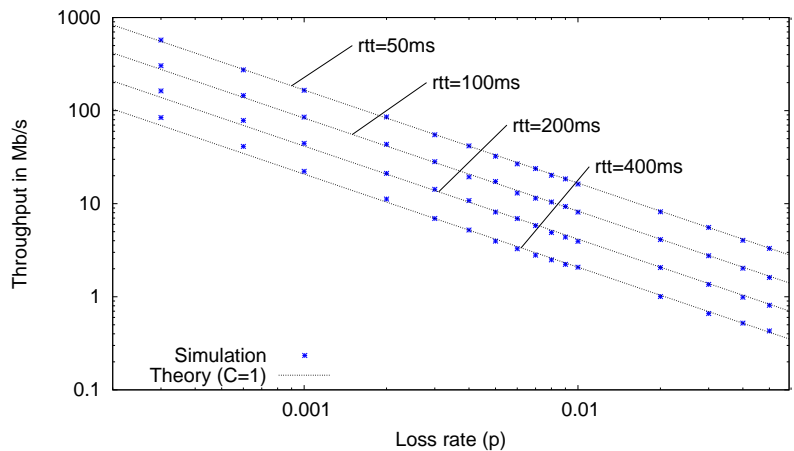}\label{result_simu_rcc_plus}}
		 \caption{Comparison of theoretical and simulated throughput of a Relentless flow} 
		 \label{result_simu_rcc}
		
\end{center}
\end{figure}

The results obtained with the ns-2 implementation of RCC are presented in Fig. \ref{result_simu_rcc_RTO}   and those obtained with the ns-2 implementation of RCC+ are presented in Fig. \ref{result_simu_rcc_plus}

In Fig. \ref{result_simu_rcc}, each point represents a simulation result with the corresponding parameters. The duration of the simulations 
is long enough to reach the steady state and the computation of the throughput is done when this steady state is reached. 
To verify the parameters of the model, simulations are done with a loss rate ranging from $0.03\%$ to $5\%$. To estimate the maximum throughput of the RCC flow, we ensure that we are not limited
by the link capacity between the source and the destination. Thus, we set this capacity to $100Gb/s$ which is much more than the maximum achievable theoretical throughput for the chosen parameters.

Fig. \ref{result_simu_rcc} shows that the model fits the simulation with $C=1$ which is the theoretical value of $C$. 
Note that we have verified inside the simulation traces that there was no retransmission timeout and confirmed that RCC+ correctly prevents RTO due to loss of retransmissions.

\section{Constant value and loss distribution}
\label{sec:constant}

In Section \ref{simulation} and in our models, we have considered uniform losses pattern. We now propose to investigate the value of the constant of the models in the case of bursty loss channel. We drive a set of experiments to evaluate $C$ and $C_{RTO}$ with an average burst size (denoted $B$) ranging from 2 to 4 following a Gilbert-Elliott channel. An important point is that RCC flow throughput still evolves in $\frac{1}{p}$ even with bursty losses. Table \ref{table_cst} gives the results obtained.

\begin{table}[htb!]
	\caption{Impact of loss model on C}
	\begin{center}
	\begin{tabular}{|c|c|c|} 
	
		\hline  \multicolumn{2}{|c|}{Loss Model}  & RCC+ \\
		\hline  \multicolumn{2}{|c|}{Uniform} & $C=1.0$ \\
		\hline  \multirow{3}{*}{Gilbert-Elliot} & $B=2$ & $C=0.90$ \\ \cline{2-3}
		   & $B=3$ & $C=0.90$ \\ \cline{2-3}
		   & $B=4$ & $C=0.90$ \\
		\hline
	\end{tabular}
	
	\label{table_cst}
	\end{center}
\end{table}

As RCC+ prevents RTO triggering, the constant remains stable making the model robust (at least up to $B=4$). However, the slight decrease of the constant value (from $1.0$ to $0.9$) might 
be explained by the increase of the average number of losses at the end of the congestion window.
Indeed, if a loss occurs at the end of the congestion window, its detection is possible only two rounds after. As the pace of the sent packets is driven by the pace of the received acknowledgements, this implies that the number of packets sent in the round following this loss is lower than the window size value. More generally, we can state that, if there are $n$ losses at the end of the congestion window, the number of packets sent in the next $TD$ is reduced by $n$.

\section{Conclusion and future work}
\label{sec:conclusion}

We have proposed a model of the Relentless Congestion Control algorithm and an ns-2 implementation (available for download at \small \url{http://personnel.isae.fr/remi-diana}\normalsize) based on SACK+. 
We confirm that RCC evolves in $\frac{1}{p}$ and our performance evaluation shows the need to enable a fair-queuing algorithm to prevent unfairness between other TCP variants.

As a next step, we propose to further assess the benefit of using RCC as a potential solution for long delay link and satellite communications. 
We also expect to use this model in a larger performance evaluation study which aims at evaluating RCC+ with various TCP variants. 

\section{Acknowledgements}

This study has been supported by funding from CNES and Thales Alenia Space.

\bibliographystyle{plain}

\end{document}